\documentclass[12pt]{article}
\def\int {\intop \limits}

\def\fnote#1{\footnote}

\begin{document}

\renewcommand \theequation{\thesection.\arabic{equation}}

\title{On propagation of high-energy photon in a medium in presence of\\
an external field}

\author{V. N. Baier
and V. M. Katkov\\
Budker Institute of Nuclear Physics,\\ Novosibirsk, 630090,
Russia}

\maketitle

\begin{abstract}
The polarization tensor is calculated which originates from
interaction of a photon with the electron-positron field. The
effects of multiple scattering of electrons and positrons in a
medium side by side with external fields is included. The corresponding 
general representation of polarization tensor is found in the form of 
two-dimensional integral. The both
effects may be essential for propagation of high-energy photons
in oriented single crystals.
\end{abstract}

\newpage

\section{Introduction}

As known, the propagation of electromagnetic wave in a medium is
defined by its dielectric tensor ${\cal E}_{ik}(\omega)$. For
relatively low frequency $\omega$ (e.g. visible light) the
dielectric tensor is defined by atomic phenomena. When the
frequency of wave is much higher the atomic frequencies, the
dielectric tensor has a form
\begin{equation}
{\cal E}_{ik}(\omega)=\delta_{ik}{\cal E}(\omega),\quad {\cal
E}(\omega)= 1-\frac{\omega_0^2}{\omega^2},\quad
\omega_0^2=\frac{4\pi n_e e^2}{m},\label{1}
\end{equation}
where $n_e$ is the number density of electrons in a medium, $m
(e)$ is the electron mass (charge), $\omega_0$ is the plasma
frequency, in any medium $\omega_0 < 100$~eV. So that for $\omega
\gg \omega_0$ an influence of atomic phenomena on propagation of
electromagnetic wave in a medium becomes small. It should be noted
that for radiation of high-energy particles the characteristic
frequency is $\gamma \omega_0,~\gamma$ is the Lorentz factor.

At very high energy the nonlinear effects of QED enter into game.
These effects are due to interaction of photon with the
electron-positron field. One of them is the polarization of the
vacuum by a photon. In the presence of an external
electromagnetic field the polarization of vacuum was considered
first in the pioneer papers \cite{S}. In the strong field this
effect can be essential for propagation of high-energy photons
\cite{N}, \cite{BKF}.

To evaluate the polarization tensor one has to consider the
amplitude of photon scattering which included the polarization
operator. We use the quasiclassical operator method \cite{BK},
\cite{BKF}, \cite{BLP}, \cite{BKS}. In this method the mentioned
amplitude is described by diagram where the virtual
electron-positron pair is first created by the initial photon
with 4-momentum $k (\omega,{\bf k})$ and polarization ${\bf e}_1$
and then annihilate into final photon with 4-momentum $k$ and
polarization ${\bf e}_2$. This corresponds to use of the
non-covariant perturbation theory where at high energies ($\omega
\gg m$) the contribution of this diagram survives only. For this
energy of photon this process occurs in a rather long time (or at
a rather long distance) known as the lifetime of the virtual state
\begin{equation}
l_f=\frac{\omega}{2q_c^2}, \label{1.1}
\end{equation}
where $q_c \geq m$ is the characteristic transverse momentum of
the process, the system $\hbar=c=1$ is used. When the virtual
electron (or positron) is moving in a medium it scatters on atoms
and changes the velocity under influence of external
electromagnetic field. The mean square of momentum transfer to the
electron from a medium and an external field on the distance $l_f$
is
\begin{eqnarray}
&& q_f^2=q_s^2+{\bf q}_F^2, \quad q_s^2=4\pi Z^2\alpha^2n_a L
l_f,\quad L \equiv L(q_c^2)=\ln
(q_c^2 a^2), \nonumber \\
&& {\bf q}_F=e{\bf F}_e l_f,\quad  {\bf F}_e={\bf E}_{\perp}+{\bf
v}\times{\bf H} \label{1.2}
\end{eqnarray}
where $\alpha=e^2=1/137$, $Z$ is the charge of nucleus, $n_a$ is
the number density of atoms in the medium,  $a$ is the screening
radius of atom, ${\bf E}_{\perp}$ is the electric field strength
transverse to the velocity of particle ${\bf v} \simeq {\bf
n}={\bf k}/\omega$, ${\bf H}$ is the magnetic field strength.

In the case of small momentum transfer $q_f \equiv \sqrt{q_f^2}
\ll m$ the influence of a medium and an external field is weak, in
this case $q_c=m$. At high energy it is possible that $q_c \geq
m$. In this case the characteristic value of the momentum transfer
(giving the main contribution into the spectral probability) is
defined by the value of $q_f$. The self-consistency condition is
\begin{eqnarray}
&& q_c^2=q_f^2=\frac{2\pi \omega Z^2\alpha^2n_a
L(q_f)}{q_f^2}+\frac{m^6 \mbox{\boldmath$\kappa$}^2}{4q_f^4} \geq
m^2,~\mbox{\boldmath$\kappa$}=\frac{\omega}{m^3}e{\bf F},
\nonumber \\
&&{\bf F}={\bf E}-{\bf n}({\bf n}{\bf E})+{\bf n}\times {\bf H},
\label{1.3}
\end{eqnarray}
here $\kappa=|\mbox{\boldmath$\kappa$}|$ is known parameter
characterising the pair production process in a homogeneous
external field ${\bf F}$. With $q_c$ increase the lifetime of the
virtual state (\ref{1.1}) decreases.

We will use the following normalization condition for the
amplitude under consideration
\begin{equation}
M=2 \omega \Delta\omega.
\label{1.4}
\end{equation}

The amplitude $M$ is the contraction of the tensor $e_j^{(i)\ast}e_k^{(f)}$
(${\bf e}^{(i)}$ and ${\bf e}^{(f)}$ are the polarization vectors
of the initial and the final photons) and the polarization tensor
$M_{jk}$. We select the basic vectors as
\begin{equation}
{\bf e}_1=\frac{{\bf F}}{|{\bf F}|},\quad {\bf e}_2={\bf n}\times{\bf e}_1.
\label{1.7}
\end{equation}
Since the tensor $M_{jk}$ is invariant under the space inversion then
in the selected basic vectors it has the diagonal form
\begin{equation}
M_{jk}=\frac{1}{2}\left[\delta_{jk}(M_{11}+M_{22})+
(e_{1j}e_{1k}-e_{2j}e_{2k})(M_{11}-M_{22}) \right]
\label{1.8}
\end{equation}
In absence of external field it is convenient to describe the
process of photon scattering using the helicity polarization
vector ${\bf e}_{\lambda}~(\lambda = \pm 1)$ connected with
momentum transfer $\mbox{\boldmath$\Delta$}$ (see Eq.(2.34) in
\cite{BK3}). In presence of external field and for
$\mbox{\boldmath$\Delta$}=0$ we choose the polarization vectors in
the following way:
\begin{equation}
{\bf e}_{\lambda}=\frac{1}{\sqrt{2}}\left({\bf e}_1 +i\lambda{\bf
e}_2 \right),\quad ({\bf e}_{\lambda}{\bf
e}_{\lambda}^{\ast})=1,\quad ({\bf e}_{\lambda} {\bf
e}_{-\lambda}^{\ast})=0,\quad {\bf e}_{\lambda}\times {\bf
n}=i\lambda{\bf e}_{\lambda}. \label{2.5}
\end{equation}

In terms of helicity amplitudes $M_{++}$ and $M_{+-}$ the tensor $M_{jk}$
and the corresponding dielectric tensor ${\cal E}_{jk}$ has a form
\begin{eqnarray}
&& M_{jk}=\delta_{jk}k_{++}^2+(e_{1j}e_{1k}-e_{2j}e_{2k})k_{+-}^2,
\nonumber \\
&& {\cal E}_{jk}=\delta_{jk}-\frac{1}{\omega^2}M_{jk},\quad
k_{++}^2 \equiv M_{++},\quad k_{+-}^2 \equiv M_{+-}
\label{1.9}
\end{eqnarray}
The polarization tensor $k_{jk}^2$  is diagonal in the basic
vectors ${\bf e}_1$ and ${\bf e}_2$ (\ref{1.7}). The corresponding
mass squared are
\begin{equation}
k_1^2 \equiv k_{11}^2=k_{++}^2+k_{+-}^2, \quad
k_2^2 \equiv k_{22}^2=k_{++}^2-k_{+-}^2
\label{2.14}
\end{equation}

The probability of pair creation by a photon with polarization
${\bf e}$ is
\begin{equation}
W_p^F({\bf e})=-\frac{1}{\omega}{\rm Im}\left[({\bf e}{\bf
e}_1)^2k_1^2+ ({\bf e}{\bf e}_2)^2k_2^2 \right]=
-\frac{1}{\omega}{\rm Im}\left[k_{++}^2+\xi_3k_{+-}^2\right],
\label{2.15}
\end{equation}
where $\xi_3$ is the Stokes' parameter. For unpolarized photon
one has
\begin{equation}
W_p^F=-\frac{1}{\omega}{\rm Im}~k_{++}^2.
\label{2.16}
\end{equation}

\section{Polarization tensor}
\setcounter{equation}{0}

The polarization of a medium by a high-energy photon in the
presence of an external field is described by the amplitudes of
photon forward scattering. The general representation of photon
scattering amplitudes without change of helicity $M_{++}=M_{--}$
and with helicity flip $M_{+-}=M_{-+}$ was obtained in
\cite{BK3} (see Eq.(3.14)). In the mentioned paper \cite{BK3} the
helicity amplitudes were normalized by the condition ${\rm
Im}M_{++}=\omega \sigma_p(\omega)$, where $\sigma_p(\omega)$ is
the total cross section of a production of electron-positron pair
by a photon with energy $\omega$. In the case under consideration
we are interested in corrections to the energy (mass) of photon
and it is convenient to use the normalization (\ref{2.15}) and
(\ref{2.16})
\begin{equation}
k_{\lambda\lambda'}^2=-2\alpha m^2\int_{0}^{\omega}
T_{\lambda\lambda'}\frac{\omega
d\varepsilon}{\varepsilon\varepsilon'}, \label{2.1}
\end{equation}
where
\begin{eqnarray}
\hspace{-10mm}&& T_{++}=T_{--},\quad T_{+-}=T_{-+},
\nonumber \\
\hspace{-10mm}&& T_{++}=\left<0|s_1\left(G^{-1}-G_0^{-1} \right)+
s_2 {\bf p}\left(G^{-1}-G_0^{-1} \right){\bf p}|0\right>,
\nonumber \\
\hspace{-10mm}&&T_{+-}=-2\left<0|s_3\left({\bf e}_{-}^{\ast}{\bf
p}\right) \left(G^{-1}-G_0^{-1} \right)\left({\bf e}_{+}{\bf
p}\right)|0\right>,~s_1=1,~s_2=\frac{\varepsilon^2+\varepsilon'^2}{\omega^2},
\nonumber \\
\hspace{-10mm}&& s_3=\frac{2\varepsilon
\varepsilon'}{\omega^2},\quad G={\cal H}+1,~{\cal H}={\bf
p}^2+V_p(\mbox{\boldmath$\varrho$}),~G_0={\bf p}^2+1,\quad \varepsilon'=
\omega-\varepsilon.
\label{2.2}
\end{eqnarray}
These expressions are similar to the expressions for
probabilities of bremsstrahlung \cite{BK1} and pair creation by
photon \cite{BK2} when the involved charged particles are
subjected to the multiple scattering in a medium. The developed
approach is given in these papers in detail. The same method was
used in \cite{BK3} for analysis of coherent scattering of a
photon in a medium. In absence of an external field one can use
the results of this paper putting the momentum transfer
$\mbox{\boldmath$\Delta$}=0$. A modification of radiative
correction under simultaneous influence of multiple scattering
and homogeneous electromagnetic field was considered recently by
authors in \cite{BK4}, where the anomalous magnetic moment of an
electron was considered. The potential
$V_p(\mbox{\boldmath$\varrho$})$ can be obtained from
Eqs.(2.7)-(2.11) of \cite{BK4} with the help of standard
substitutions $\omega \rightarrow -\omega, \varepsilon
\rightarrow -\varepsilon$:
\begin{equation}
V_p(\mbox{\boldmath$\varrho$})=-iV(\mbox{\boldmath$\varrho$})+
2\mbox{\boldmath$\kappa'$}\mbox{\boldmath$\varrho$},\quad
\mbox{\boldmath$\kappa'$}=\frac{\varepsilon\varepsilon'}{\omega^2}
\mbox{\boldmath$\kappa$}, \label{2.3}
\end{equation}
where $\mbox{\boldmath$\kappa$}$ and ${\bf F}$ are defined in
Eq.(\ref{1.3}). The potential $V(\mbox{\boldmath$\varrho$})$ has
the form \cite{BK1}, \cite{BK2}:
\begin{eqnarray}
&&
\hspace{-4mm}V(\mbox{\boldmath$\varrho$})=Q\mbox{\boldmath$\varrho$}^2
\left(L_1+\ln \frac{4}{\mbox{\boldmath$\varrho$}^2}-2C \right),
\quad Q=\frac{2\pi Z^2\alpha^2\varepsilon \varepsilon'
n_a}{m^4\omega},\quad L_1=\ln \frac{a_{s2}^2}{\lambda_c^2},
\nonumber \\
&& \hspace{-4mm}\frac{a_{s2}}{\lambda_c}=183Z^{-1/3}{\rm
e}^{-f},\quad
f=f(Z\alpha)=(Z\alpha)^2\sum_{k=1}^{\infty}\frac{1}{k(k^2+(Z\alpha)^2)},
\label{2.4}
\end{eqnarray}
here $C=0.577216 \ldots$ is Euler's constant, $n_a$ is defined in
(\ref{1.2}),  $\mbox{\boldmath$\varrho$}$ is the coordinate in
the two-dimensional space measured in the Compton wavelength
$\lambda_c$, which is conjugate to the space of the transverse
momentum transfers measured in the electron mass $m$.

We split the potential $V_p(\mbox{\boldmath$\varrho$})$ in the
same way as in Eqs.(2.9)-(2.11) of \cite{BK4}:
\begin{eqnarray}
&&
V_p(\mbox{\boldmath$\varrho$})=V_{pF}(\mbox{\boldmath$\varrho$})
-iv(\mbox{\boldmath$\varrho$}),\quad
V_{pF}(\mbox{\boldmath$\varrho$})=-iV_c(\mbox{\boldmath$\varrho$})+
2\mbox{\boldmath$\kappa'$}\mbox{\boldmath$\varrho$}, \quad
V_c(\mbox{\boldmath$\varrho$})= q\mbox{\boldmath$\varrho$}^2,
\nonumber \\
&& q=QL_c \quad, L_c \equiv L(\varrho_c) =\ln
\frac{a_{s2}^2}{\lambda_c^2 \varrho_c^2},\quad
v(\mbox{\boldmath$\varrho$})=-\frac{q\mbox{\boldmath$\varrho$}^2}{L_c}
\left(\ln \frac{\mbox{\boldmath$\varrho$}^2}{4\varrho_c^2}+2C
\right).
\label{2.6}
\end{eqnarray}
Here the parameter $\varrho_c$ is defined by the set of equations
(compare with Eq.(2.20) of \cite{BK4}):
\begin{eqnarray}
&& \varrho_c =1\quad {\rm for}\quad
4(\mbox{\boldmath$\kappa'$}^2+QL_1) \leq 1;
\nonumber \\
&& 4\varrho_c^4 \left(\mbox{\boldmath$\kappa'$}^2\varrho_c^2+QL_c
\right)=1 \quad {\rm for} \quad 4(\mbox{\boldmath$\kappa'$}^2+
QL_1) \geq 1,
\label{2.7}
\end{eqnarray}
According to this splitting and taking into account the addition
to the potential (\ref{2.3}) we present the propagators in
Eq.(\ref{2.2}) as
\begin{equation}
G^{-1}-G_0^{-1} = G^{-1} - G_{pF}^{-1}+ G_{pF}^{-1} -G_0^{-1},
\label{2.8}
\end{equation}
where
\begin{equation}
 G_{pF}= {\cal H}_{pF}+1,\quad G=G_{pF}-iv, \quad
{\cal H}_{pF} = {\bf p}^2+V_{pF}.
\label{2.9}
\end{equation}
The representation of the propagator $G^{-1}$ permits to carry out
its decomposition over the "perturbation" $v$
\begin{equation}
G^{-1} - G_{pF}^{-1} = G_{pF}^{-1}~ iv~ G_{pF}^{-1} + G_{pF}^{-1}
~iv~ G_{pF}^{-1}~iv~G_{pF}^{-1} + \ldots
\label{2.10}
\end{equation}
The procedure of matrix elements calculation in this decomposition
was formulated in \cite{BK1}, see Eqs.(2.30), (2.31). Here the
basic matrix element is \newline
$<\mbox{\boldmath$\varrho$}_1|G_{pF}^{-1}|\mbox{\boldmath$\varrho$}_2>$.
This matrix element can be obtained from Eq.(2.15) in \cite{BK4}
making the substitutions:$\omega \rightarrow -\omega, \varepsilon \rightarrow
-\varepsilon, \mbox{\boldmath$\chi$}/u \rightarrow\mbox{\boldmath$\kappa'$}$
\begin{eqnarray}
\hspace{-12mm}&&
<\mbox{\boldmath$\varrho$}_1|G_{pF}^{-1}|\mbox{\boldmath$\varrho$}_2>
=i\int_{0}^{\infty}dt \exp(-it) K_{pF}(\mbox{\boldmath$\varrho$}_1,
\mbox{\boldmath$\varrho$}_2, t),
\nonumber \\
\hspace{-12mm}&& K_{pF}(\mbox{\boldmath$\varrho$}_1,
\mbox{\boldmath$\varrho$}_2, t)=K_c(\mbox{\boldmath$\varrho$}_1,
\mbox{\boldmath$\varrho$}_2,
t)K_{\kappa}(\mbox{\boldmath$\varrho$}_1,
\mbox{\boldmath$\varrho$}_2, t),
\label{2.11}
\end{eqnarray}
where
\begin{eqnarray}
\hspace{-16mm}&&
K_c(\mbox{\boldmath$\varrho$}_1, \mbox{\boldmath$\varrho$}_2, t)=
\frac{\nu}{4\pi i \sinh \nu t} \exp \left\{ \frac{i\nu}{4}
\left[ (\mbox{\boldmath$\varrho$}_1^2+\mbox{\boldmath$\varrho$}_2^2)
\coth \nu t - \frac{2}{\sinh \nu t}
\mbox{\boldmath$\varrho$}_1\mbox{\boldmath$\varrho$}_2\right] \right\},
\nonumber \\
\hspace{-16mm}&&K_{\kappa}(\mbox{\boldmath$\varrho$}_1,
\mbox{\boldmath$\varrho$}_2,
t)=\exp\left[-\frac{4i\mbox{\boldmath$\kappa'$}^2t}{\nu^2}\left(
1-\frac{2}{\nu t}\tanh \frac{\nu
t}{2}\right)-\frac{2i}{\nu}\mbox{\boldmath$\kappa'$}
(\mbox{\boldmath$\varrho$}_1+
\mbox{\boldmath$\varrho$}_2)\tanh \frac{\nu t}{2}\right].
\label{2.12}
\end{eqnarray}
where $\nu=2\sqrt{iq}$,~ $q$ is defined in (\ref{2.6}).

In the present paper we restrict ourselves to the main term in
the decomposition (\ref{2.8}). This means that result will have
the logarithmic accuracy over the scattering (but not over an
external field). The matrix elements entering into the mass
correction have in the used approximation the following form
\begin{eqnarray}
&& M_1= \left<0|G_{pF}^{-1}-G_0^{-1} |0\right>
=\frac{1}{4\pi}\int_{0}^{\infty}\exp(-it)\left(\frac{\nu}{\sinh
\nu t}\varphi_p -\frac{1}{t} \right)dt,
\nonumber \\
&& M_2= \left<0|{\bf p}(G_{pF}^{-1}-G_0^{-1}){\bf p} |0\right>
=-\frac{1}{4\pi}\int_{0}^{\infty}\exp(-it)\Bigg[\frac{\nu \varphi_p}{\sinh
\nu t} \Bigg(\frac{4\kappa'^2}{\nu^2}\tanh^2\frac{\nu t}{2}
\nonumber \\
&& +\frac{i\nu}{\sinh \nu t} \Bigg) -\frac{i}{t^2}
\Bigg]dt,
\nonumber \\
&& M_3=\left<0|({\bf e}_{-}^{\ast}{\bf p})(G_{pF}^{-1}-
G_0^{-1})({\bf e}_{+}{\bf p}) |0\right> =-\frac{\kappa'^2}{2\pi
\nu} \int_{0}^{\infty}\exp(-it)\frac{\varphi_p}{\sinh \nu t}
\tanh^2\frac{\nu t}{2}dt,
\nonumber \\
&& \varphi_p \equiv \varphi(\mbox{\boldmath$\kappa'$}, \nu, t)
=\exp\left[-\frac{4i\kappa'^2t}{\nu^2}\left( 1-\frac{2}{\nu
t}\tanh\frac{\nu t}{2} \right) \right].
\label{2.13}
\end{eqnarray}

Integrating by parts the term containing $1/\sinh^2 \nu t$
together with the subtraction term in expression for $M_2$ we find
\begin{equation}
M_2=-\frac{\nu}{4\pi}\int_{0}^{\infty}\exp(-it)\varphi_p
\tanh\frac{\nu t}{2}\left(\frac{4\kappa'^2}{\nu^2}+1\right)dt -M_1
\label{2.17}
\end{equation}
Substituting the found expressions for $M_1$ and $M_2$ into
formula (\ref{2.2}) and then into Eq.(\ref{2.1}) we obtain the
general expressions for photon masses squared under simultaneous
influence of multiple scattering in a medium and an external
electromagnetic field
\begin{eqnarray}
&& k_{++}^2=\frac{\alpha m^2}{2\pi}\int_{0}^{\omega} \frac{\omega
d\varepsilon}{\varepsilon\varepsilon'}\int_{0}^{\infty}e^{-it}
\nonumber \\
&& \times \left[s_2 \nu \varphi_p \tanh\frac{\nu
t}{2}\left(\frac{4\kappa'^2}{\nu^2}+1\right)-s_3\left(\frac{\nu}{\sinh
\nu t}\varphi_p -\frac{1}{t} \right)\right]dt \label{2.18}
\end{eqnarray}
For $k_{+-}^2$ we found respectively
\begin{equation}
k_{+-}^2= -\frac{2\alpha m^2}{\pi}\int_{0}^{\omega} \frac{\omega
d\varepsilon}{\varepsilon\varepsilon'}s_3
\kappa'^2\int_{0}^{\infty}e^{-it}\frac{\varphi_p}{\nu \sinh \nu
t}\tanh^2\frac{\nu t}{2}dt \label{2.19}
\end{equation}

In the absence of external field ($\mbox{\boldmath$\kappa'$}=0,~\varphi_p=1$)
we have
\begin{eqnarray}
&& k_{++}^2=\frac{\alpha m^2}{2\pi}\int_{0}^{\omega} \frac{\omega
d\varepsilon}{\varepsilon\varepsilon'}\int_{0}^{\infty}e^{-it}
\left[s_2 \nu \tanh\frac{\nu t}{2}
-\frac{s_3}{ 2t}\left(\frac{\nu}{\sinh \nu t}
-\frac{1}{t} \right)\right]dt
\nonumber \\
&& = -\frac{\alpha m^2}{2\pi}\int_{0}^{\omega} \frac{\omega
d\varepsilon}{\varepsilon\varepsilon'}\left[s_1\left(
\ln p-\psi\left(p+\frac{1}{2} \right) \right) +s_2
\left(\psi(p)-\ln p +\frac{1}{2p} \right)\right],
\nonumber \\
&& k_{+-}^2=0,
\label{2.20a}
\end{eqnarray}
where $p=i/2\nu,~ \psi(p)$ is the logarithmic derivative of the
gamma function. Subsisting the result obtained into formula
(\ref{2.16}) we have the probability of pair creation which
agrees with formula (2.10) in \cite{BK2}.

In the absence of multiple scattering $(\nu \rightarrow 0)$ we get
\begin{eqnarray}
&& k_{++}^2=\frac{\alpha m^2}{\pi}\int_{0}^{\omega} \frac{\omega
d\varepsilon}{\varepsilon\varepsilon'}\int_{0}^{\infty}e^{-it}
\left[s_2 \kappa'^2 t \varphi_F -\frac{s_3}{ 2t}\left(\varphi_F
-1 \right)\right]dt,
\nonumber \\
&& k_{+-}^2= -\frac{\alpha m^2}{2\pi}\int_{0}^{\omega}
\frac{\omega d\varepsilon}{\varepsilon\varepsilon'}s_3
\kappa'^2\int_{0}^{\infty}e^{-it}\varphi_F tdt,
\label{2.20}
\end{eqnarray}
where
\begin{equation}
\varphi_F=\exp\left(-i\frac{\kappa'^2 t^3}{3}\right)
\label{2.21}
\end{equation}
For this case the expressions for $k_{1,2}^2$ (\ref{2.14}) after
substitution results of (\ref{2.20}) agree with masses squared of
photon calculated in an external electromagnetic field (see
\cite{BKF} and references therein).

We will analyse now the results obtained in different limiting cases.
In the case when the both characteristic parameters are small ($\nu_1^2
=4QL_1 \ll 1,~\kappa \ll 1$), the main terms
of decomposition of the functions
$k^2$ are the sum of independent corrections to the photon mass squared
both on account of the multiple scattering and an external field
\begin{eqnarray}
\hspace{-16mm}&& k_{++}^2=\frac{\alpha m^2}{\pi}
\Bigg[-i\frac{7}{9}\frac{\omega}{\omega_e}
\left(1-\frac{1}{21L_1} \right)+\frac{59}{225}
\left(\frac{\omega}{\omega_e}\right)^2
\nonumber \\
\hspace{-16mm}&& -i\frac{3\sqrt{3}\pi}{16\sqrt{2}}\frac{\omega}{\omega_F}
\exp\left(-\frac{8\omega_F}{3\omega}\right)-\frac{11}{90}
\left(\frac{\omega}{\omega_F}\right)^2 \Bigg],
\nonumber \\
\hspace{-16mm}&& k_{+-}^2= \frac{\alpha m^2}{\pi} \Bigg[
i\frac{\sqrt{3}\pi}{16\sqrt{2}} \frac{\omega}{\omega_F}
\exp\left(-\frac{8\omega_F}{3\omega}\right)+\frac{1}{30}
\left(\frac{\omega}{\omega_F}\right)^2 \Bigg];\quad \omega \ll
\omega_e,\quad \omega \ll \omega_F. \label{2.22}
\end{eqnarray}
Here the notations are introduced:
\begin{equation}
\omega_e=\frac{m}{2\pi Z^2 \alpha^2 \lambda_c^3 n_a L_1},\quad
\omega_F=m \frac{H_0}{|{\bf F}|}, \quad H_0=\frac{m^2}{e}=4.41
\cdot 10^{13}{\rm Oe}.
\label{2.23}
\end{equation}
For used notations see Eqs.(\ref{1.3}), (\ref{2.4}). The
correction $\propto 1/L_1$ follows from the first term of
decomposition in Eq.(\ref{2.10}). In gold the value $\omega_e$ is
$\omega_e=10.5$~TeV, this is the typical value for the heavy elements.

In Fig. the functions Re~$k_{++}^2$(curve 2) and Im~$k_{++}^2$ (curve 1)
are given for the case when the influence of a medium is taken into account
only (Eq. (\ref{2.20a})). The both curves are normalized to the 
asymptotics given
by Eq. (\ref{2.22}) in the limit $\omega_F \rightarrow \infty$.

In the case when the influence of an external field is small comparing
with effect of a medium $\omega/\omega_F \ll (1+\omega/\omega_e)^{3/4}$
we can present the expression for $k^2$ as
\begin{equation}
k^2=k^2(\omega_c)+\Delta k^2(\omega_F).
\label{2.24}
\end{equation}
Here $k^2(\omega_c)$ is the photon mass squared under influence
of a medium only (in absence of an external field $\kappa'=0$) is
given by Eq.(\ref{2.20a}), where the function $L_c$ is defined in
Eqs.(\ref{2.6}) and (\ref{2.7})
\begin{equation}
\nu^2=4iq=i\frac{4\varepsilon \varepsilon'}{\omega^2}
\frac{\omega}{\omega_c},\quad \omega_c=\omega_e\frac{L_1}{L_c},
\quad L_c \simeq L_1 \left[1+\frac{1}{2L_1}\ln \left(1
+\frac{\omega}{\omega_e} \right) \right]. \label{2.26}
\end{equation}
Retaining the main terms of decomposition over
$\mbox{\boldmath$\kappa'$}^2$ in Eqs.(\ref{2.18}) and
(\ref{2.19}) we find the corrections $\Delta k^2 (\omega_F)$:
\begin{eqnarray}
&& \Delta k_{++}^2(\omega_F)=-i\frac{\alpha m^2}{2\pi}
\frac{\omega \omega_c}{\omega_F^2} \int_{0}^{\omega} \frac{
d\varepsilon}{\omega}\int_{0}^{\infty}\exp \left(-i\frac{z}{\nu} \right)
\nonumber \\
&& \times \left\{s_2 \left[\tanh\frac{z}{2}
\left(1-i\frac{z}{\nu}\right)+\frac{2i}{\nu}\tanh^2\frac{z}{2}\right]+
\frac{is_3}{\nu \sinh z}\left( z -2 \tanh\frac{z}{2}
\right)\right\}dz,
\nonumber \\
&& \Delta k_{+-}^2(\omega_F)=i\frac{\alpha m^2}{2\pi}
\frac{\omega \omega_c}{\omega_F^2} \int_{0}^{\omega} \frac{
d\varepsilon}{\omega}s_3\int_{0}^{\infty}
\exp \left(-i\frac{z}{\nu} \right)
\frac{\tanh^2\frac{z}{2}}{\sinh z}dz.
\label{2.27}
\end{eqnarray}

In the limit $\omega \gg \omega_e$ one can use the asymptotics found
in Appendix A in \cite{BK2} for calculation of the photon mass squared
$k_{++}^2(\omega_c)$:
\begin{eqnarray}
&& k_{++}^2(\omega_c)=-\frac{2\alpha m^2}{\pi}
\Pi(a),\quad a^2=i\frac{\omega_c}{\omega},
\nonumber \\
&& \Pi(a) \simeq -\frac{3\pi}{8a} +\left(2\ln 2-\frac{1}{2}
\right)\ln \frac{1}{a}+ A -\frac{\pi^3 a}{48},
\nonumber \\
&& A=\ln2 (2\ln 2 -1) + \frac{1}{2}(1+C)-2\sum_{n=1}^{\infty}
\frac{(-1)^n}{n}\ln n = 0.736629. \label{2.28}
\end{eqnarray}
It should be mentioned that in the limit $\omega_F \rightarrow \infty$ 
this formulas gives $k_{++}^2$ at $\omega /\omega_e = 10$ within
accuracy better than $7\%$ and at $\omega /\omega_e = 100$ within accuracy
better than $0.3\%$.

The corrections $\Delta k^2 (\omega_F)$ due to action of external
field in the same limit $\omega \gg \omega_e$ are
\begin{equation}
\Delta k_{++}^2(\omega_F) \simeq -i\frac{2\alpha m^2}{3\pi}
\frac{\omega \omega_c}{\omega_F^2} (1-\ln 2), \quad
\Delta k_{+-}^2(\omega_F) \simeq i\frac{\alpha m^2}{12\pi}
\frac{\omega \omega_c}{\omega_F^2}.
\label{2.29}
\end{equation}

In the case when the principal effect is due to an external field,
the main contribution into the photon mass squared is given by
Eq.(\ref{2.20}) and corrections connected with influence of
multiple scattering are
\begin{eqnarray}
&& k^2=k^2(\omega_F)+ \Delta k^2(\omega_c),
\nonumber \\
&& \Delta k_{++}^2(\omega_c)=\frac{\alpha m^2}{15\pi}
\frac{\omega}{\omega_c} \int_{0}^{\omega} \frac{
d\varepsilon}{\omega}\Bigg\{\int_{0}^{\infty} \exp
\left[-it\left(1+\frac{1}{3}\kappa'^2t^2 \right)\right]
\nonumber \\
&& \times \left[s_2 \left(8it +3 +\frac{2}{\kappa'^2}
\right)+s_3\left(3it+\frac{1}{\kappa'^2} \right)\right]dt+
\frac{i}{\kappa'^2}(2 s_2+s_3)\Bigg\},
\nonumber \\
&& \Delta k_{+-}^2(\omega_c)=-\frac{2\alpha m^2}{15\pi}
\frac{\omega}{\omega_c} \int_{0}^{\omega} \frac{
d\varepsilon}{\omega}s_3\Bigg[\int_{0}^{\infty} \exp
\left[-it\left(1+\frac{1}{3}\kappa'^2t^2 \right)\right]
\left(2it-3+\frac{1}{2\kappa'^2} \right)
\nonumber \\
&&+\frac{i}{2\kappa'^2}\Bigg].
\label{2.30}
\end{eqnarray}
Here the function $L_c$ is determined in Eqs.(\ref{2.6}) and (\ref{2.7})
at $Q=0$
\begin{equation}
L_c \simeq L_1 \left[1+\frac{2}{3L_1}\ln \left(1
+\frac{\omega}{\omega_F} \right) \right].
\label{2.31}
\end{equation}
To obtain expressions Eq.(\ref{2.30}) we performed decomposition
in formulas (\ref{2.18}) and (\ref{2.19}) and integrated by parts
the terms with high powers of $t$ in integrals over $t$. Note,
that using Eq.(\ref{2.16}) we can obtain from Eq.(\ref{2.30}) the
correction to the probability of pair photoproduction due to
effect of weak multiple scattering. The result agrees with
Eq.(7.136) in \cite{BKS}.

In the limiting case $\omega \gg \omega_F$ we have from Eq.(\ref{2.20})
\begin{equation}
k_{++}^2(\omega_F) \simeq \frac{5\alpha m^2}{7\pi}
\displaystyle{e^{-i\frac{\pi}{3}}}\frac{\Gamma^3(2/3)}{\Gamma(1/3)}
\left(\frac{3 \omega}{\omega_F}\right)^{2/3}, \quad
k_{+-}^2(\omega_F) =- \frac{1}{5} k_{++}^2(\omega_F). \label{2.32}
\end{equation}
For corrections originating from the multiple scattering we have from
Eq.(\ref{2.30}) the main term of decomposition
\begin{equation}
\Delta k_{++}^2(\omega_c) \simeq \frac{4\alpha m^2}{75\pi}
e^{-i\frac{\pi}{6}}\frac{\Gamma^3(1/3)}{\Gamma(2/3)}
\left(\frac{3 \omega}{\omega_c}\right)^{1/3}
\left(\frac{\omega_F}{\omega_c}\right)^{2/3}, \quad \Delta
k_{+-}^2(\omega_c) = \frac{1}{2}\Delta k_{++}^2(\omega_c).
\label{2.33}
\end{equation}

\section{Conclusion}
\setcounter{equation}{0}

It is curious that in the scope of the used method (see
e.g.\cite{BK3}) it is possible to find many of obtained in
previous section results (up to numerical coefficients) basing on
very simple form of the amplitude of photon scattering
\begin{equation}
M \sim \frac{\alpha}{\pi} q^2,
\label{1.5}
\end{equation}
where $q^2$ is defined by the equation
\begin{equation}
q^2=-i\frac{2\pi \omega Z^2 \alpha^2 n_a
L(q^2+m^2)}{q^2+m^2}-\frac{m^6
\mbox{\boldmath$\kappa$}^2}{4(q^2+m^2)^2}. \label{1.6}
\end{equation}

In the case $q_f^2 \ll m^2~(|q|^2 \simeq q_f^2)$, see
Eqs.(\ref{1.2}) and (\ref{1.3}), the imaginary part of the photon
scattering amplitude in forward direction (Eqs.(\ref{1.5}) and
(\ref{1.6}))is defined by the value $q_s^2$ while the real part of
this amplitude is a sum of the correction $\sim q_s^4/m^2$ and the
momentum transfer due to action of a field $q_F^2$ (compare with
(\ref{2.22}))
\begin{equation}
M \sim \frac{\alpha}{\pi}
m^2\left[-i\frac{q_s^2(m)}{m^2}+\frac{q_s^4(m)}{m^4}-\frac{\kappa^2}{4}\right].
\label{a.1}
\end{equation}

At strong multiple scattering ($q_s^2 \gg m^2$) and in the case
$q_F^2 \ll q_s^2~(\kappa^2 m^6 \ll q_s^6)$ the value $q_s^2 \simeq
q_f^2$ is defined by Eqs.(\ref{1.2}) and (\ref{1.3})
\begin{equation}
q_s^4(q_s)=2\pi \omega Z^2 \alpha^2 n_a L(q_s^2),
\label{a.2}
\end{equation}
and the photon scattering amplitude (\ref{1.5}) is (compare with
Eqs.(\ref{2.28}) and (\ref{2.29}))
\begin{equation}
M \sim \frac{\alpha}{\pi}
\left[e^{-i\frac{\pi}{4}}q_s^2(q_s)-i\frac{m^6\kappa^2}{4
q_s^4(q_s) }\right].
 \label{a.3}
\end{equation}

When the main effect is the action of the field $\kappa \gg m$ and
in the case $q_s^2(q_F^2) \ll q_F^2~(\kappa^2 m^6 \gg q_s^4
q_F^2)$ the value $q_F^2 \sim q_f^2$ we have (see (\ref{1.3}))
\begin{equation}
q_F^6(q_F)=\frac{m^6 \kappa^2}{4}, \label{a.4}
\end{equation}
and the photon scattering amplitude (\ref{1.5}) is (compare with
Eqs.(\ref{2.32}) and (\ref{2.33}))
\begin{equation}
M \sim \frac{\alpha}{\pi}
\left[e^{-i\frac{\pi}{3}}q_F^2(q_F)+e^{-i\frac{\pi}{6}}q_s^2(q_F)\right].
 \label{a.5}
\end{equation}

It should be noted that beginning with some photon energy
$\omega=\omega_b$ the radiative correction to the value Re~${\cal
E}_{jk}$  in the absence of a field
($\mbox{\boldmath$\kappa$}=0$) becomes larger than
$\omega_0^2/\omega^2$ (see (\ref{1})). Let us estimate $\omega_b$.
According to formulas (\ref{1.5}) and (\ref{1.6}) this effect
originate at values $q^2 \ll m^2$. Because of this for estimate
Re~$M$ (\ref{1.5}) it is necessary to take into account the next
term in decomposition over $q^2$ in (\ref{1.6}) at
$\mbox{\boldmath$\kappa$}=0$. We find
\begin{eqnarray}
&& {\rm Re}~M_s \sim \frac{\alpha}{\pi}\frac{q_s^4(m)}{m^2},\quad
-\frac{1}{\omega}{\rm Im}~M_s \sim
\frac{\alpha}{\pi}\frac{q_s^2}{\omega}=\frac{1}{L_{rad}},
\nonumber \\
&& \frac{{\rm Re}~M_s}{\omega_0^2} \sim \frac{\pi}{\alpha}
\frac{\omega^2}{\omega_0^2}\frac{\lambda_c^2}{L_{rad}^2}, \quad
\omega_b \sim \sqrt{\frac{\alpha}{\pi}}\frac{L_{rad}}{\lambda_c}
\omega_0.\label{1.10}
\end{eqnarray}
For gold one obtains $\omega_b \sim 40$~GeV.

A propagation of high-energy photons in oriented single crystals
is one of interesting applications of the result obtained above.
In this case we have both the dense matter with strong effect of
multiple scattering and high fields of crystal axes or planes. As
known, the Landau-Pomeranchuk-Migdal (LPM) effect (influence of
multiple scattering on processes of bremsstrahlung and pair
creation by a photon) is most pronounced in the heavy elements.
The same is valid for the process under consideration. Let the
high-energy photon incident on crystal. The angle of incidence is
small and such that the distance from axis
$\mbox{\boldmath$\varrho$}$ (or the distance from plane $x$) can
be considered as a constant on the formation length of process
(the constant field approximation is applicable, see Sections
12,15 in \cite{BKS}).

For orientation of a crystal along an axis the ratio of density of
atoms in the vicinity of axis $n(\mbox{\boldmath$\varrho$})$ to
the mean density $n_a$ is
\begin{equation}
\xi_{ax}(\mbox{\boldmath$\varrho$})=\frac{n(\mbox{\boldmath$\varrho$})}{n_a}
=\frac{\exp(-\varrho^2/2u_1^2)}{2\pi u_1^2d n_a},
\label{3.1}
\end{equation}
where $u_1$ is the amplitude of thermal vibrations of atoms, $d$
is the mean distance between atoms which form the axis. This
ratio is maximal at $\mbox{\boldmath$\varrho$}=0$.  For numerical
estimates we use for definiteness the tungsten single crystal. For
the axis $<111>$ in W the ratio $\xi_{ax}(0)$=370 at the room
temperature ($T=293~K$) and $\xi_{ax}(0)$=1020 at $T=77~K$. The
effect of multiple scattering becomes strong at characteristic
photon energy $\omega_e(n_a) \simeq 11~$TeV and this value is
inversely proportional to the density. So we have that
$\omega_e(\varrho=0) \simeq 30~$GeV at $T=293~K$ and
$\omega_e(\varrho=0) \simeq 11~$GeV at $T=77~K$. It should be
noted that within logarithmic accuracy we neglect by relatively
small variation of $L_1$ due to substitution the screening radius
$a_s^2$ by the value $2u_1^2$.

It is useful to compare these estimates with known threshold
energies $\omega_t$ at which the probability of pair creation in
the field of axis is equal to the probability of the
Bethe-Heitler mechanism, see Table 12.1 in \cite{BKS}. For photon
energy  $\omega \geq \omega_t$ the process of pair creation in
the field of axis dominates. In W crystal for $<111>$ axis
$\omega_t=22~$GeV at T=293~K and $\omega_t=13~$GeV at T=77~K. It
is seen that for these energies the ratio $\omega/\omega_e$ which
characterize the strength the LPM effect is of the order of unity.
At $\omega \sim \omega_t$ the maximal value of the parameter
$\kappa(\mbox{\boldmath$\varrho$})$ which determines the
probability of pair creation by a photon in a field is also of
the order of unity (at $\varrho \simeq u_1$, see the mentioned
Table). So we reach the conclusion that at some energy (for axial
orientation of crystal) all the discussed effects are essential
simultaneously. The analysis in this situation will be published
elsewhere. For example, to calculate the influence of the field
of axis on the polarization tensor one have to average the
general formula (\ref{2.18}) over all values of
$\mbox{\boldmath$\varrho$}$ (this is integration over
$d^2\varrho$ with the weight $n_{\perp}$, where $n_{\perp}=n_a d$
is the density of axis in the perpendicular to them plane).

At planar orientation of crystal the ratio of the density of
atoms in the vicinity of plane $n(x)$ to the mean density $n_a$ is
\begin{equation}
\xi_{pl}(x)=\frac{n(x)}{n_a} =\frac{\exp(-x^2/2u_1^2)d_{p
l}}{\sqrt{2\pi} u_1}, \label{3.2}
\end{equation}
where $d_{pl}$ is the distance between planes. For the plane
$(110)$ in W crystal at $T=293~K$ one has $\xi_{pl}(0) \sim 18$
and effective value $\omega_e \simeq 600~$GeV. This value is
$\sim 2.5$ higher than the threshold energy $\omega_t$ for this
plane (see Table 15.1 in \cite{BKS}). Let us note that in both
axial and planar cases we made estimations for the maximal value
of $\xi(0)$.

In the crystals where the atomic number $Z$ is not very high (Ge,
Si, C) the ratio $\omega_t/\omega_e(0)$ is smaller than unity. So
for $\omega \geq \omega_t$ one can use Eqs.(\ref{2.20}) and
(\ref{2.30}) while for $\omega \ll \omega_t$ the formula
(\ref{2.22}) is applicable in which along with known results (see
e.g. \cite{BKS}) there is the new term in Re~$k_{++}^2$ which is
proportional to $(\omega/\omega_e)^2$.

{\bf Acknowledgements}
\vspace{0.2cm}
This work
was supported in part by the Russian Fund of Basic Research under Grant
00-02-18007.

\newpage

{\bf Figure caption}

\vspace{15mm}The functions Re~$k_{++}^2$(curve 2) and Im~$k_{++}^2$ (curve 1)
versus the photon energy taken in units
$\omega_e$ (because of this the curves are universal) for the case when the 
influence of a medium is taken into account
only (Eq.(\ref{2.20a})). The both curves are normalized to 
the asymptotics given
by Eq.(\ref{2.22}) in the limit $\omega_F \rightarrow \infty$.

\end{document}